\documentclass[runningheads]{svmult}

\usepackage{makeidx}   
\usepackage{graphicx}  
\usepackage{subeqnar}  
\usepackage{multicol}  
\usepackage{physprbb}  
\makeindex             

\begin{document}
\title*{Broad-band Modelling of GRB Afterglows}
\toctitle{Broad-band, Self-consistent Modelling of GRB Afterglows}
%
%
\titlerunning{Modelling of Gamma-ray Burst Afterglows}
%
\author{Edo Berger\inst{1}
\and Re'em Sari\inst{1}
\and Dale Frail\inst{2}
\and Shri Kulkarni\inst{1}}
\authorrunning{Edo Berger et al.}
%
%
\institute{Division of Physics, Mathematics \&\ Astronomy 105-24,
         California Institute of Technology, Pasadena, CA 91125, USA
\and National Radio Astronomy Observatory, P. O. Box O, Socorro, NM 87801, USA}

\maketitle              

\begin{abstract}
\index{abstract} Observations of GRB afterglows ranging from radio to X-ray 
frequencies generate large data sets. Careful analysis of these broad-band
data can give us insight into the nature of the GRB progenitor
population by yielding such information like the total energy of the
burst, the geometry of the fireball and the type of environment into
which the GRB explodes. We illustrate, by example, how global,
self-consistent fits are a robust approach for characterizing the
afterglow emission. This approach allows a relatively simple
comparison of different models and a way to determine the strengths
and weaknesses of these models, since all are treated self-consistently. 
Here we quantify the main differences between the broad-band, 
self-consistent approach and the traditional approach, using 
GRB\thinspace000301C and GRB\thinspace970508 as test cases. 
\end{abstract}

\section{Introduction}
The quest for an understanding of GRB and afterglow physics, as well as
the parameters that characterize the burst has recently led us to a new 
approach to the modelling of afterglow data.  In principle, by modelling 
the afterglow data it is possible to extract the five parameters 
characterizing the synchrotron spectrum ($\nu_a$, $\nu_m$, $\nu_c$,
$p$, and $F_{\nu_m}$) from which we can calculate the burst energy, the 
ambient medium density, and the fractions of energy in the magnetic fields 
and electrons~\cite{SPN98}.  At the same time, with accurate modelling it 
is possible to distinguish between the different models of afterglow 
emission, i.e. ISM vs. wind, and spherical vs. collimated 
outflow~\cite{BSF00}~\cite{SPH99}~\cite{R99}~\cite{CL99}~\cite{CL00}.

\section{The Shortcomings of the Traditional Approach}
Since the discovery of afterglow emission from GRBs in the late 1990s, 
the general approach to afterglow modelling has consisted of the following 
steps~\cite{WG99}.  The data set collected for a particular burst was 
broken up into lightcurves and spectra, which were fitted separately.  
The spectra were modelled using the broken synchrotron spectrum in order 
to extract the value of $p$, and possibly the break frequencies.  The 
lightcurves were each fitted separately to solve for the temporal decay 
slopes, $\alpha_i$, which were then compared for consistency, and in the 
optical band to extract any host galaxy extinction.  The temporal decay 
slopes were also used to distinguish between the different models
of afterglow emission, and breaks were used to infer the existence of a jet 
geometry.  This approach has several serious drawbacks:
\vspace{-0.08in}
\begin{itemize}
\item Only a few data points are modelled at a time, and the uncertainty in 
the derived parameters is large.  
\item The deduced model parameters and power-law indices are not always 
physically meaningful (e.g. can give $\epsilon_B>1$).
\item Since this approach employs the broken synchrotron power-law, the 
modelling of lightcurves and spectra near the break frequencies is inaccurate.  
\item It is extremely difficult to account for the changes in the spectrum 
and time dependences when the order of the break frequencies changes.
\end{itemize}

\section{The Advantages of the Broad-band Approach}
Our approach attempts to remedy the aforementioned problems, and in addition 
to clearly identify the present shortcomings of afterglow studies.  The 
procedure we use in modelling the data is significantly different.  We use a 
broad-band data set ranging from radio to X-rays and fit it simultaneously~\cite{BSF00}.  
We therefore give equal weight to all data points, and do not disregard scattered 
data points, which in the traditional approach are useless.  Our approach also tests 
a complete model with all its different early and late time variations, including 
the transition to the sub-relativistic phase.  It is therefore self-consistent since 
it does not include or exclude any assumptions and constraints that are part of 
the complete model.  With this approach we gain the following advantages:
\vspace{-0.075in}
\begin{itemize}
\item All data points are used simultaneously since the model includes both the 
temporal and frequency dependence of each parameter.
\item We use the Granot, Piran and Sari smoothed synchrotron 
spectrum~\cite{GPS99a}~\cite{GPS99b}, which is a much more accurate and realistic 
representation of the actual data.  
\item We can easily include all special cases of the spectral and temporal evolution; 
therefore, any significant deviation of the data from the predicted models can be 
interpreted as a possibly new phenomenon 
(e.g. GRB\thinspace000301C~\cite{BSF00}~\cite{GLS00}).
\item We can easily extract the values of the burst energy, ambient density and 
fractions of energy in the magnetic fields and electrons. 
\item We can directly determine which model (e.g. ISM vs. Wind) gives the most 
accurate description of the data using a simple $\chi^2$ statistic.
\end{itemize}

\section{Conclusion}
The study of GRB afterglows and the extraction of the burst characteristics 
from the observations can be severly limited if a narrow-band approach is used.  
The problems of this traditional approach to modelling are numerous, but they 
can be easily solved if a broad-band, self-consistent approach is used instead.  
We have shown that the overall behavior of the afterglow emission can be easily 
studied within this approach, that the correct emission model can be unambiguously 
identified if the data set is large enough, and that the parameters characterizing 
the burst (e.g. energy, ambient density) can be easily solved for.

\begin{figure}
\begin{center}
\mbox{{\includegraphics[width=.32\textwidth]{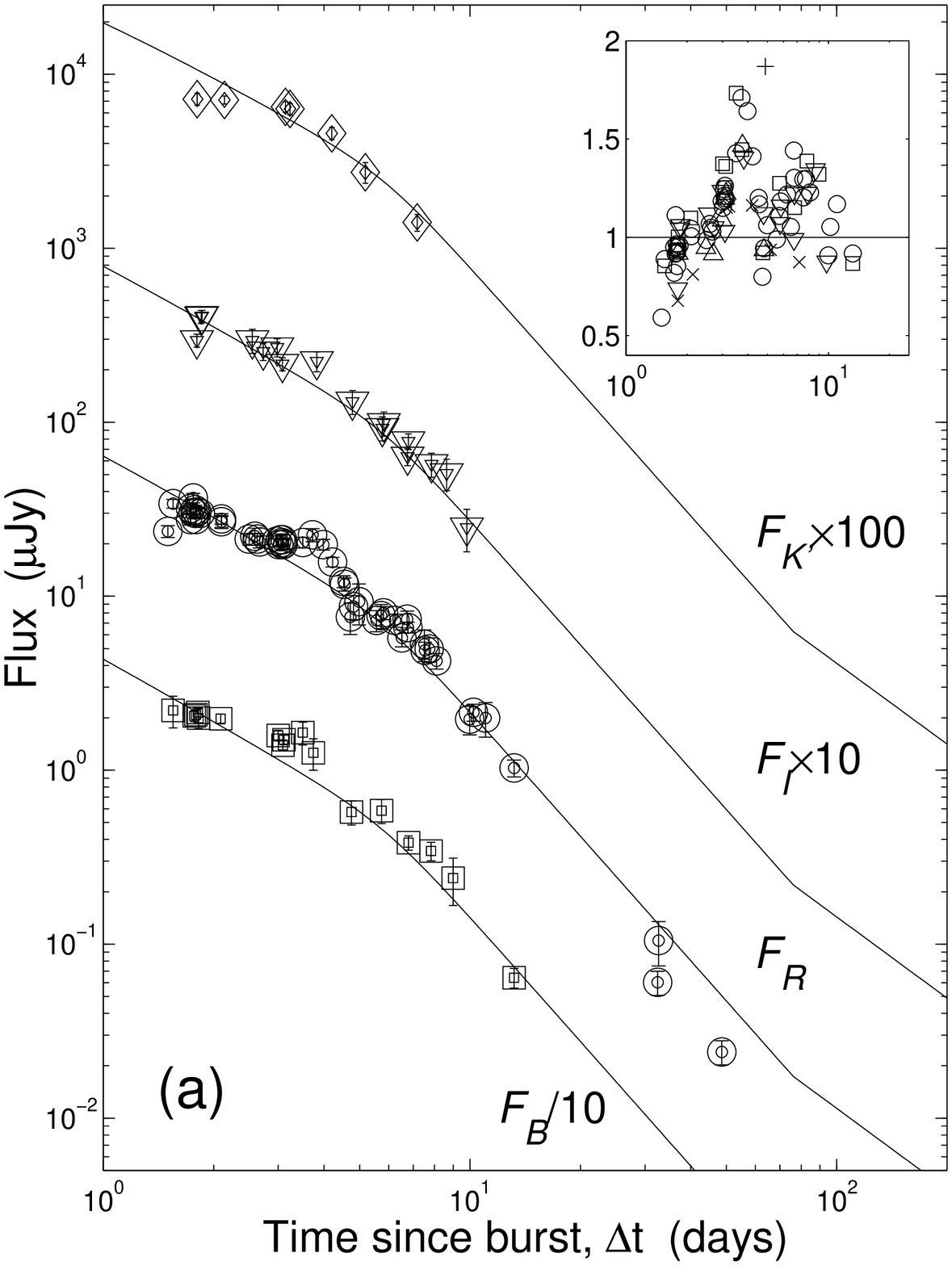}}\hfill
{\includegraphics[width=.32\textwidth]{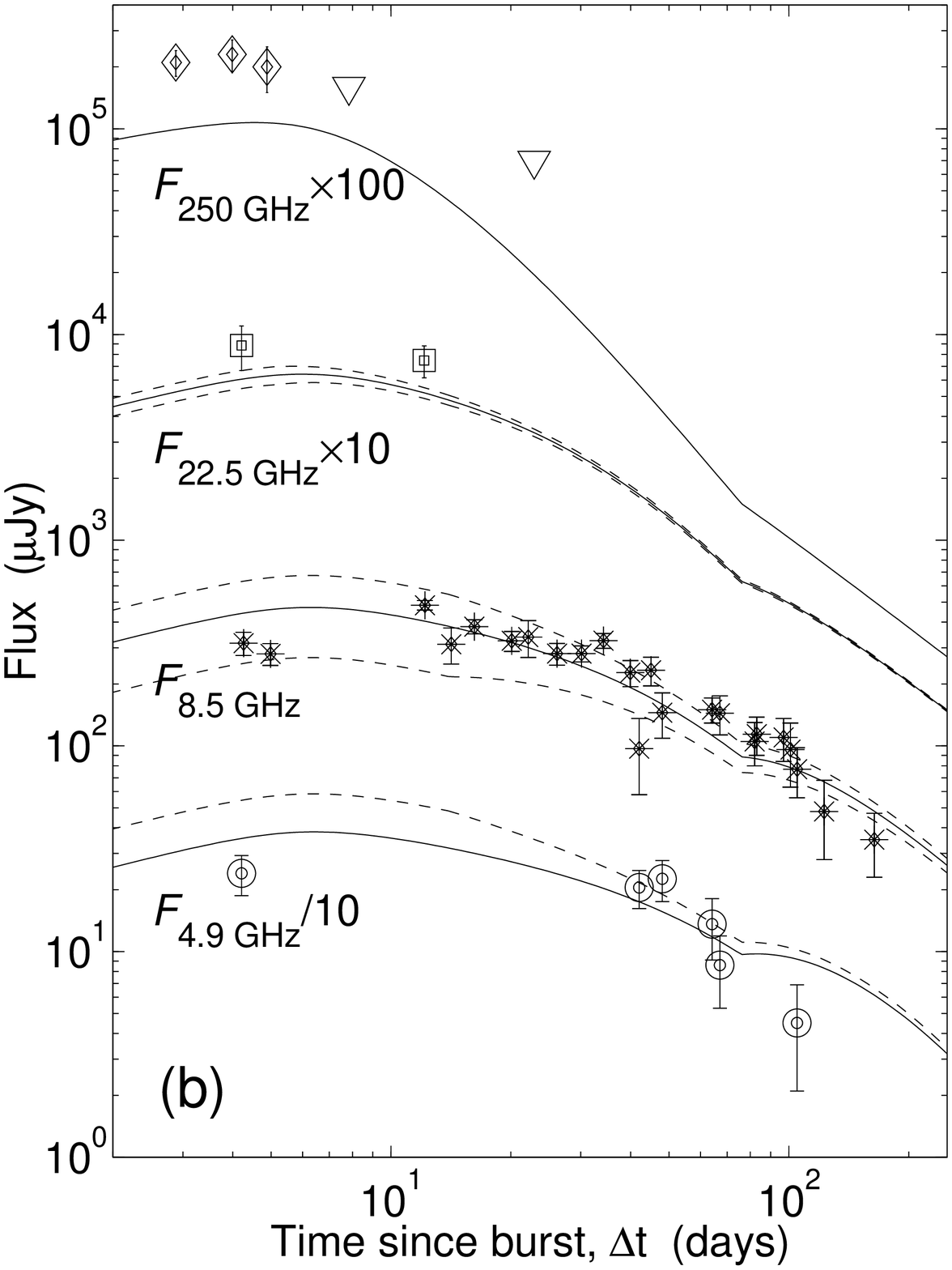}}
{\includegraphics[width=.32\textwidth]{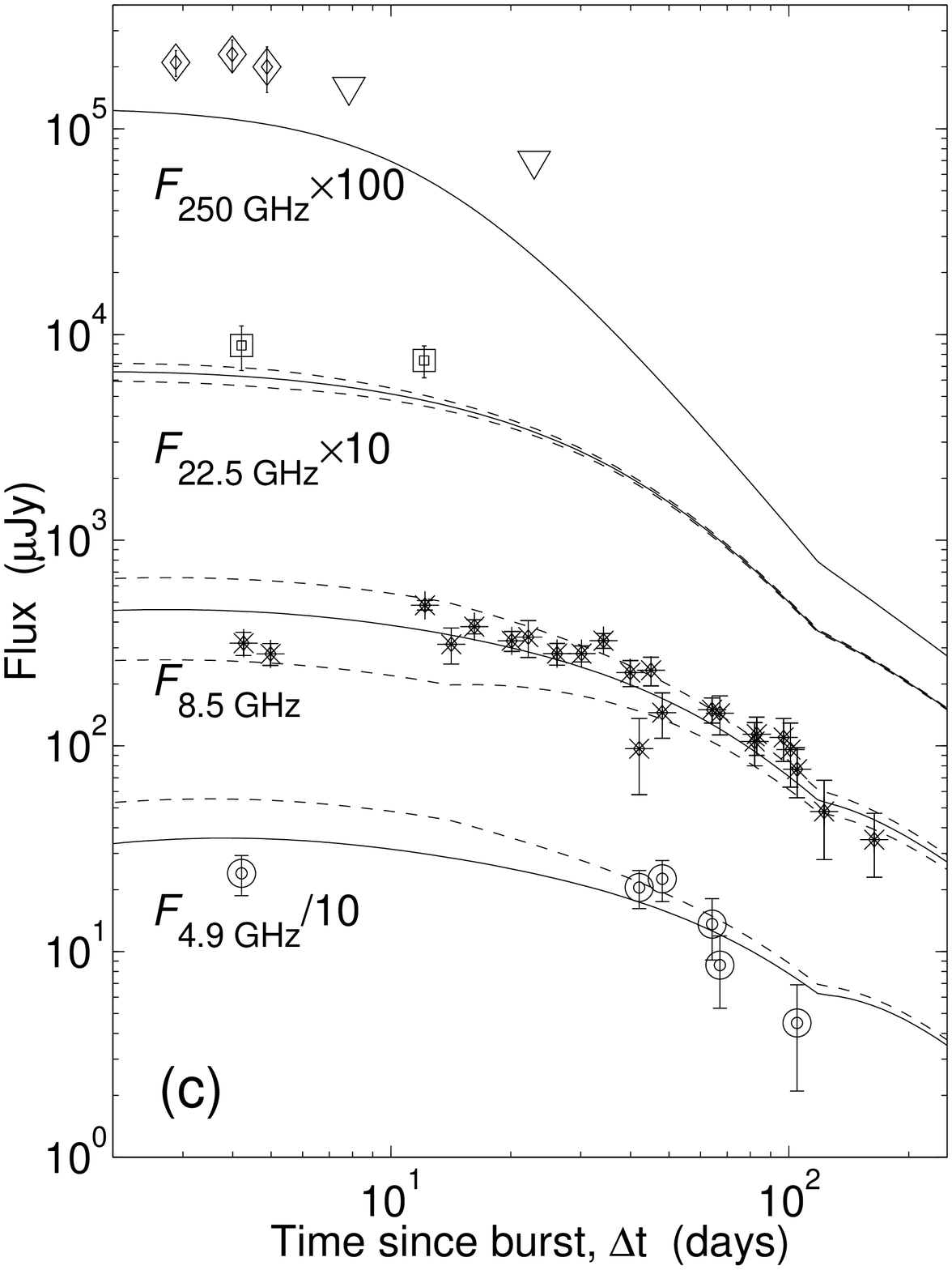}}}
\end{center}
\vspace{-0.1in}\caption[]{\small (a) Optical and (b) radio lightcurves of 
GRB\thinspace000301C for the ISM+jet model and (c) the wind+jet model. The 
dashed lines indicate flux variation due to scintillation.  The models include 
a jet break and a non-relativistic phase.  The insert shows the achromatic bump 
which was only evident as a result of the global fitting~\cite{BSF00}.}
\end{figure}

\begin{figure}
\begin{center}
\mbox{{\includegraphics[width=.32\textwidth]{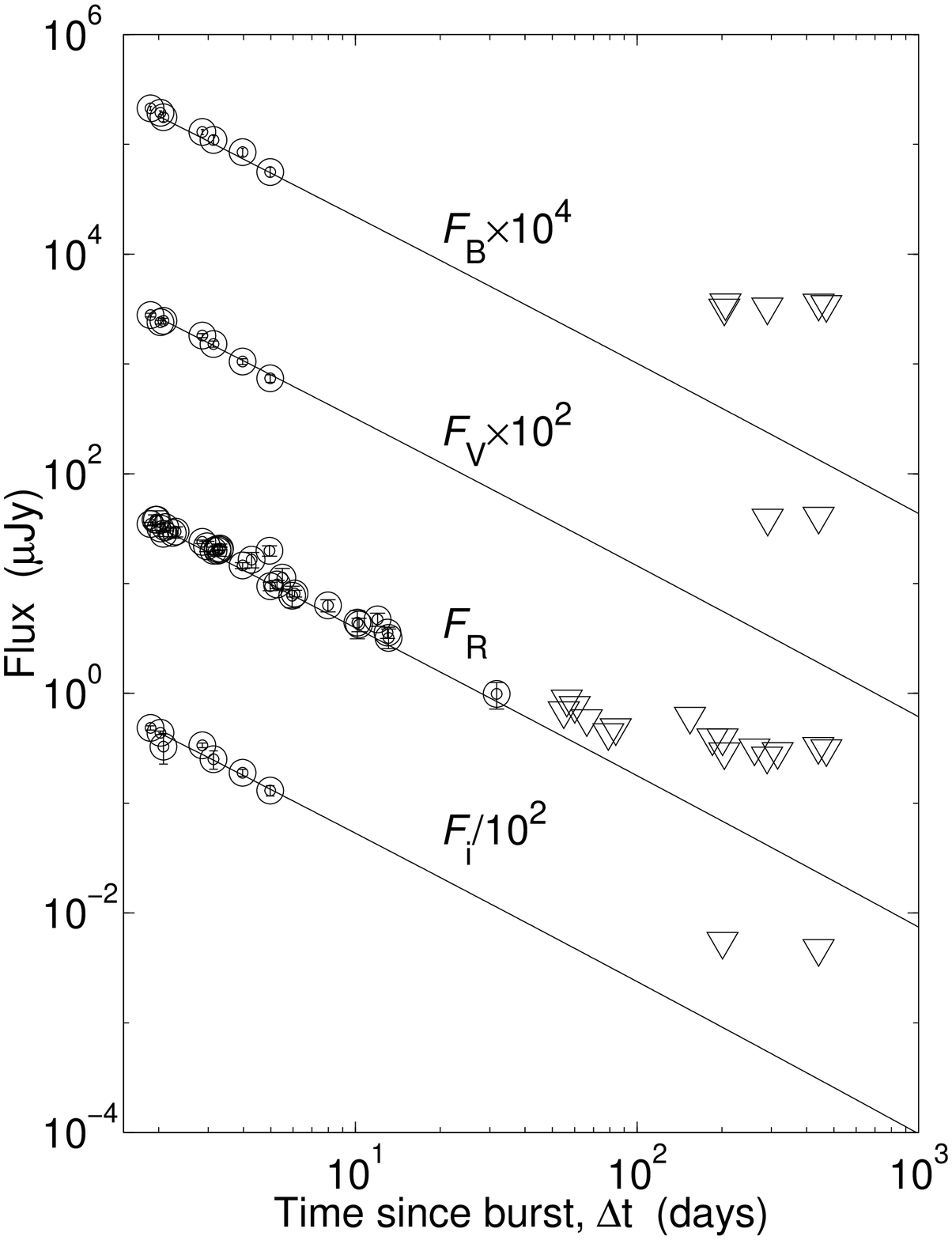}}\hfill
{\includegraphics[width=.32\textwidth]{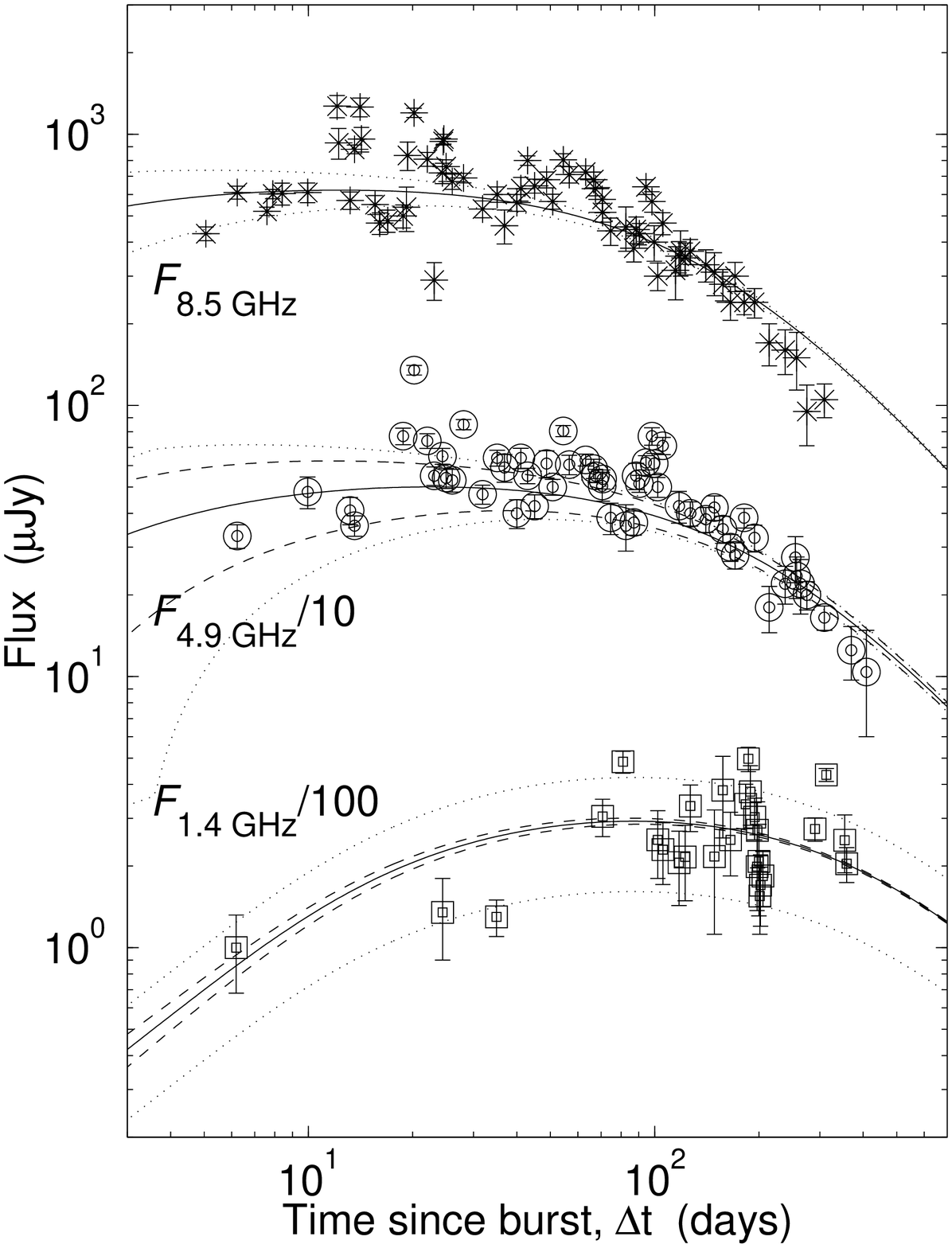}}
{\includegraphics[width=.32\textwidth]{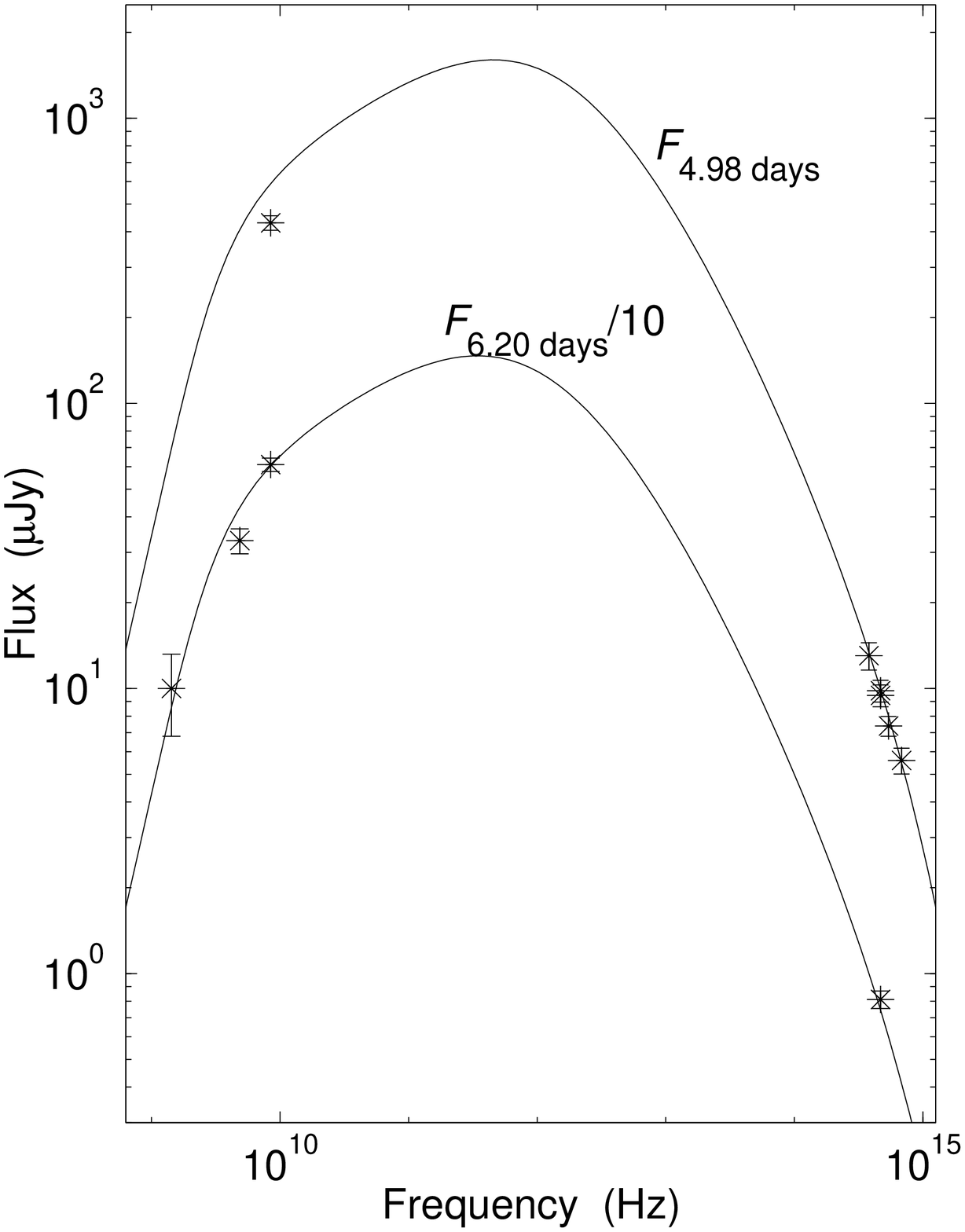}}}
\end{center}
\vspace{-0.1in}\caption[]{\small Optical and radio lightcurves and spectra of 
GRB\thinspace970508 for the wind model.  The Modelling includes the effect of 
host galaxy extinction and host flux density.  Upper limits in the optical 
indicate measurements in which the host galaxy flux dominates.}
\end{figure}

%

\end{document}